\documentclass{PoS}
\usepackage{amsmath}
\newcommand{\Comments}[1]{}

\newcommand{\nn}{\nonumber}

\newcommand{\VEV}[1]{\left\langle{#1}\right\rangle}
\newcommand{\chibar}{{\bar{\chi}}}
\newcommand{\Psfig}[2]{\includegraphics[width=#1]{Figs/#2}}
\newcommand{\Feff}{{\cal F}_\mathrm{eff}}
\newcommand{\Vq}{{\cal V}_{q}}

\newcommand{\Od}{{\cal O}}
\newcommand{\Vp}{V^{+}}
\newcommand{\Vm}{V^{-}}
\newcommand{\bsig}{b_\sigma}

\newcommand{\Fint}{{\cal D}}
\newcommand{\wt}{{\omega_\tau}}

\newcommand{\bt}{{\beta_\tau}}

\newcommand{\Seff}{S_\mathrm{eff}}

\newcommand{\Zp}{Z_{+}}
\newcommand{\Zm}{Z_{-}}

\newcommand{\tilmu}{\tilde{\mu}}

\newcommand{\psiT}{\psi_{\tau}}

\newcommand{\psiTT}{\psi_{\tau\tau}}
\newcommand{\psiTS}{\psi_{\tau s}}
\newcommand{\psiSS}{\psi_{ss}}
\newcommand{\psibarT}{\bar{\psi}_{\tau}}

\newcommand{\psibarTT}{\bar{\psi}_{\tau\tau}}
\newcommand{\psibarTS}{\bar{\psi}_{\tau s}}
\newcommand{\psibarSS}{\bar{\psi}_{ss}}

\newcommand{\bs}{\beta_s}
\newcommand{\bts}{\beta_{\tau s}}
\newcommand{\btt}{\beta_{\tau\tau}}
\newcommand{\bss}{\beta_{ss}}
\newcommand{\bsigp}{b'_\sigma}
\newcommand{\mq}{m_q}

\title{Phase diagram and critical point evolution\\
in NLO and NNLO strong coupling lattice QCD}

\ShortTitle{Phase diagram and critical point evolution in NLO and NNLO SC-LQCD}

\author{\speaker{Akira Ohnishi}\\
        Yukawa Institute for Theoretical Physics,
	Kyoto University, Kyoto 606-8502, Japan\\
        E-mail: \email{ohnishi@yukawa.kyoto-u.ac.jp}}

\author{Kohtaroh Miura\\
        Yukawa Institute for Theoretical Physics,
	Kyoto University, Kyoto 606-8502, Japan}

\author{Takashi Z. Nakano\\
	Department of Physics, Faculty of Science,
	Kyoto University, Kyoto 606-8502, Japan}

\author{Noboru Kawamoto\\
	Department of Physics, Faculty of Science,
	Hokkaido University, Sapporo 060-0810, Japan}

\abstract{
We investigate the chiral phase transition
in the strong coupling lattice QCD (SC-LQCD)
at finite temperature and density with finite coupling effects.
We adopt one species of staggered fermion,
and develop an analytic formulation based 
on strong coupling and cluster expansions.
We derive the effective potential as
a function of two order parameters, 
the chiral condensate $\sigma$ and the vector potential $\wt$,
in a self-consistent treatment of
the next-to-leading order (NLO) and the next-to-next-to-leading order (NNLO)
effective action terms.
Finite coupling effects lead to modifications of
quark mass, chemical potential and
the quark wave function renormalization factor.
Finite coupling effects suppress
the critical temperature at $\mu=0$ ($T_{c,\mu=0}$),
while critical temperature at $T=0$ ($\mu_{c,T=0}$) is not affected much.
NNLO corrections does not significantly affect $T_{c,\mu=0}$ and $\mu_{c,T=0}$,
but the phase diagram shape including the position of the critical point
is sensitive to the NNLO effects.
Partially chiral restored matter is found to exist in NLO and NNLO SC-LQCD.
}

\FullConference{The XXVII International Symposium on Lattice Field Theory - LAT2009\\
		 July 26-31 2009\\
		 Peking University, Beijing, China}

\begin{document}



\section{Introduction}
\label{sec:Intro}

Understanding the quantum chromodynamics (QCD) phase diagram is
one of the most interesting problems in quark and hadron physics.
Since the lattice QCD Monte-Carlo (MC) simulation has a sign problem
at finite chemical potential $\mu$~\cite{sign-problem},
it is necessary to invoke some approximations in QCD
or to apply effective models in order to elucidate the whole phase boundary,
including the position of the critical point~\cite{CEP}.
The strong-coupling lattice QCD (SC-LQCD) is
one of the most instructive approximations especially to investigate
the phase structure at finite temperature $T$ and chemical potential $\mu$.
Based on the successes of SC-LQCD in pure Yang-Mills 
theory~\cite{YM}, chiral transition at finite $T$ and $\mu$
has been investigated at strong coupling~\cite{SCL,KlubergStern,Bilic,SCL-PD}.
For example,
the phase diagram structure has been predicted in the strong coupling limit
(SCL)~\cite{Bilic,SCL-PD},
and it is recently confirmed qualitatively in MC simulations%
~\cite{deForcrand:2009dh}
based on the monomer-dimer-polymer (MDP) formalism~\cite{MDP}.
In order to make a step forward towards the true phase diagram,
it is necessary to develop the formalism to include the plaquette effects
both in MC simulations and SC-LQCD.

In this proceedings, we evaluate the effective potential in SC-LQCD 
with one species of unrooted staggered fermion at finite $T$ and $\mu$
including the next-to-leading
(NLO, $\mathcal{O}(1/g^2)$)~\cite{Miura_QY,PDevol}
and the next-to-next-to-leading (NNLO, $\mathcal{O}(1/g^4)$)~\cite{Nakano}
effects,
and investigate the coupling dependence
of the phase diagram and the critical point.

\section{Effective Potential in NNLO SC-LQCD}
\label{Sec:Feff}

The effective potential $\Feff$
is obtained in two steps in a finite $T$ treatment of SC-LQCD.
We first integrate out the spatial links $U_j$
and obtain the effective action $\Seff$.
In the next step
we integrate out the quark field $\chi$ and the temporal links $U_0$,
and obtain the effective potential $\Feff$.
We define the effective action and potential ($\Seff$ and $\Feff$)
on a lattice with spatial (temporal) size $L(N_\tau)$
at chemical potential $\mu$ as,
\begin{align}
&\exp\left[-S_\mathrm{eff}(\chi,\chibar,U_0)\right]
= \int \mathcal{D} U_j~e^{-S_\mathrm{LQCD}}
= \int \mathcal{D}U_j~e^{-S_F-S_G}
= e^{-S_\mathrm{SCL}}
\ \big\langle e^{-S_G} \big\rangle
\label{Eq:Seff}
\ ,\\
&\exp\left[-L^d N_\tau \Feff\right]
= \int \Fint[\chi,\chibar,U_0]~\exp\left[-\Seff\right]
\ ,\\
&S_F
	=\frac12\sum_x\left[\Vp_x(\mu)-\Vm_x(\mu)\right]
	+m_0\sum_x M_x
	+\frac12\sum_{x,j}
		\eta_{j,x}
		\left[
			\chibar_xU_{j,x}\chi_{x+\hat{j}}
			-\chibar_{x+\hat{j}}U^\dagger_{j,x}\chi_x
		\right]
\ ,\label{eq:SFs}\\
&S_G
	= \frac{2N_c}{g^2} \sum_P
		\Bigl[
1-\frac{1}{2N_c}
\bigl[U_P+U_P^{\dagger}\bigr]
\Bigr]
\ ,\label{eq:SG}
\end{align}
where $d=3$, $m_0$ and $\eta_{j,x}=(-1)^{x_0+\cdots +x_{j-1}}$
are the spatial dimension, the bare quark mass and the staggered phase factor,
respectively,
$S_\mathrm{LQCD}=S_F+S_G$ is the lattice QCD action,
and $U_P$ denotes the trace of a plaquette $P$.
Mesonic composites are defined as
$M_x = \chibar_x \chi_x$,
$\Vp_x =\bar{\chi}_x e^{\mu} U_{0,x} \chi_{x + \hat{0}}$
and
$\Vm_x = \chibar_{x+\hat{0}} e^{-\mu}U_{0,x}^{\dagger} \chi_x$.

In order to perform the strong coupling expansion at finite $T$ systematically,
the cumulant (or coupled cluster) expansion is indispensable.
By using the cumulant expansion.
the expectation value of $\exp[-S_G]$ is found to be,
\begin{align}
\big\langle e^{-S_G}\big\rangle
\equiv&\frac{1}{e^{-S_\mathrm{SCL}}}
\int\mathcal{D}U_j~e^{-S_F}~e^{-S_G}
=\exp\left[
\sum_{n=1}^{\infty}\frac{(-1)^n}{n!}\big\langle S_G^n \big\rangle_c\right]
\ ,\quad
e^{-S_\mathrm{SCL}}
=\int\mathcal{D}U_j~e^{-S_{F}}
\ .
\end{align}
The bracket $\langle\cdots\rangle_c$ is called a cumulant
and shows the connected diagram contributions,
e.g. $\VEV{S_G^2}_c=\VEV{S_G^2}-\VEV{S_G}^2$.
We find that the sum in the exponent corresponds to the strong coupling
expansion of the effective action,
\begin{align}
\Seff
=&S_\mathrm{SCL}-\sum_{n=1}^{\infty}
\frac{(-1)^n}{n!}
\big\langle S_G^n \big\rangle_c
=S_\mathrm{SCL}+\Delta S_\mathrm{NLO}+\Delta S_\mathrm{NNLO}
+\Od(1/g^6, 1/\sqrt{d})
\ .\label{Eq:SeffB}
\end{align}
The $n$-th term in the sum is proportional to $1/g^{2n}$,
and we can identify
$n=1$ and $n=2$ terms as NLO and NNLO corrections
($\Delta S_\mathrm{NLO}$ and $\Delta S_\mathrm{NNLO}$),
and $S_\mathrm{SCL}$ shows the SCL effective action.

\begin{figure}[bt]
\begin{center}
\Psfig{15.2cm}{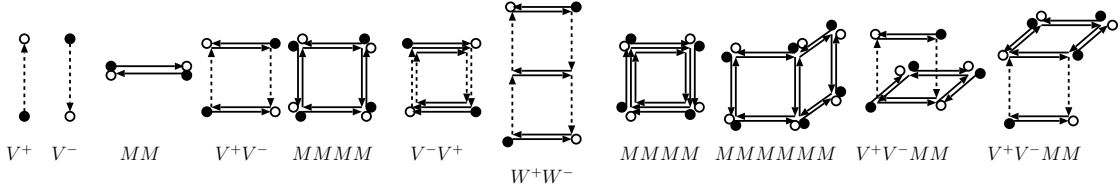}
\caption{
Diagrams contributing to the SCL, NLO and NNLO effective action terms.
Quarks (anti-quarks) are shown in open (filled) circles,
spatial (temporal) link variables are represented by solid (dotted) arrows.
and plaquettes are represented by squares.
Note that for NLO and NNLO diagrams, we also consider the hermite conjugate
of these diagrams.
}\label{Fig:diagram}
\end{center}
\end{figure}

We consider zero, one, and two-connected plaquette configurations,
and integrate out $U_j$ after putting fermionic elements
($\chibar U\chi$ and $\chibar U^\dagger\chi$).
In the leading order of the $1/d$ expansion~\cite{KlubergStern},
we find that the diagrams shown in Fig.~\ref{Fig:diagram} 
contribute to the effective action,
\begin{align}
&\Seff
= \frac12 \sum_x (V^+_x - V^-_x) - \frac{b_\sigma}{2d}\sum_{x,j>0} [MM]_{j,x}
\label{Eq:SeffSCL}
\\
&+ \frac12\,\frac {\bt}{2d} \sum_{x, j>0} 
  [V^+V^- + V^-V^+]_{j,x}
- \frac12\,\frac{\beta_s}{d(d-1)} \sum_{x, j>0,k>0,k\not=j} 
  [MMMM]_{jk,x}
\label{Eq:SeffNLO}
\\
&-\frac {\beta_{\tau\tau}}{2d} \sum_{x,j>0} 
  [W^+W^- + W^-W^+]_{j,x}
-\frac {\beta_{ss}}{4d(d-1)(d-2)}
     \sum_{\substack{x,\,j>0,\,|k|>0,\,|l|>0\\|k|\neq j,\,|l|\neq j,\,|l|\neq|k|}}
     [MMMM]_{jk,x}
     [MM]_{j,x+\hat{l}}
\nn\\
&+ \frac {\beta_{\tau s}}{8d(d-1)} \sum_{x, j > 0, \mid k \mid \neq j}
  [V^+V^- + V^-V^+]_{j,x}
     \left( [MM]_{j,x+\hat{k}} + [MM]_{j,x+\hat{k}+\hat{0}} \right)
\label{Eq:SeffNNLO}
\ ,
\end{align}
We have introduced a short-hand notation,
$[AB]_{j,x} = A_xB_{x+\hat{j}}$, 
$[ABCD]_{jk,x}= A_xB_{x+\hat{j}}C_{x+\hat{j}+\hat{k}}D_{x + \hat{k}}$.
Additional mesonic composites appears as
$W_{x}^+=e^{2\mu}\bar{\chi}_x U_{0,x}U_{0,x+\hat{0}} \chi_{x+2\hat{0}}$
and
$W_{x}^-=e^{-2\mu} \bar{\chi}_{x+2\hat{0}}U_{0,x+\hat{0}}^{\dagger}U_{0,x}^{\dagger}\chi_x$.
The coefficients are defined as,
$
b_\sigma=d/2N_c\ ,\ \ 
\bt=d(1+1/2g^2)/N_c^2g^2\ ,\ \ 
\beta_s=d(d-1)(1+1/2g^2)/8N_c^4g^2\ ,\ \ 
\beta_{\tau\tau}=d/2N_c^3g^4 \ ,\ \ 
\beta_{ss}=d(d-1)(d-2)/16N_c^7g^4\ ,\ \ 
\beta_{\tau s}=d(d-1)/2N_c^5g^4$.

We can convert the above effective action
in a spatially local and bi-linear form of quark fields
by using the extended Hubbard-Stratonovich (EHS) transformation%
~\cite{Miura_QY,PDevol}.
The fermion part of the transformed effective action is found to be
in the form of,
\begin{align}
\Seff^{(F)}=\frac12\sum_x(C \Vp_x-\bar{C}\Vm_x)
+\sum_x m M_x -\btt\sum_x(\psiTT W^+_x+\psibarTT W^-_x)
\ ,
\end{align}
where $\psiTT$ and $\psibarTT$ are the auxiliary fields
and $C$, $\bar{C}$, $m$ contain auxiliary field contributions.
We further reduce this effective action by introducing a gluonic dressed
fermion,
\begin{align}
\chi_x' = \chi_x - 2\btt\psiTT e^{\mu} U_{0,x}\,\chi_{x+\hat{0}}/C
\ ,\quad
\chibar_x' = \chibar_x + 2\btt\psibarTT \chibar_{x+\hat{0}} e^{-\mu} U^\dagger_{0,x}/\bar{C}
\ .
\end{align}
With $\chi'$ and $\chibar'$, we can absorb
the next-to-nearest neighbor (NNN) interaction terms via $W^\pm$
up to $\mathcal{O}(1/g^6)$
in the modification of mass and the coefficients of $V^\pm$,
$m$, $C$ and $\bar{C}$.

We now find that the NLO and NNLO corrections
lead to the coefficient modification
of $V^+$, $V^-$ and $M$ in the fermionic effective action.
These modifications are interpreted as
the modification of the wave function renormalization factor $Z_\chi$,
quark mass $\mq$ and chemical potential $\tilmu$.
The effective potential $\Feff$ is obtained in a similar way to that in SCL,
and is found to be,
\begin{align}
\Feff
=&\Feff^{(X)}+{\cal V}_q(\mq;\tilde{\mu},T) - N_c\log \sqrt{\Zp\Zm}
\label{Eq:Feff}
\ ,\\
\Feff^{(X)}
=& \frac12 \bsigp \,\sigma^2 +\frac12\bt'\,\psibarT\psiT
+\frac12\bs'\varphi_s^2 +\btt\psibarTT\psiTT
+\bss\psibarSS\psiSS +\frac12\bts\psibarTS\psiTS
\ ,\\
\mq=& \frac{m'}{\sqrt{\Zp\Zm}} \ ,\quad
m'= \bsigp \sigma + m_0 - \btt(\psibarTT+\psiTT)\ ,\quad
\tilde{\mu}
=\mu-\log\sqrt{\Zp / \Zm}
\label{Eq:modification} 
\ ,\\
\Vq
=&-T\log\left[
	    \frac {\sinh[(N_c+1)E_q/T]}{\sinh[E_q/T]} +2\cosh(N_c\tilmu/T)
	\right]
\label{Eq:FeffMq}
\ ,\quad
E_q(\mq)=\mathrm{arcsinh}\,(\mq)
\ ,\\
\bt'=&\bt+\bts\psi_{\tau s}\ ,\quad
\bs'=\bs+2\bss\psibarSS
\ ,\quad
\bsigp= \bsig
	+2\bs'\varphi_s
	+2\bss\psi_{ss}+2\bts\bar{\psi}_{\tau s}
\ ,\\
\Zp=& 1+\bt'\psibarT+4\btt m'\psibarTT
\ ,\quad
\Zm= 1+\bt'\psiT   +4\btt m'\psiTT
\ .
\end{align}

The auxiliary fields introduced during the bosonization procedure
have to satisfy the stationary condition,
$\partial\Feff/\partial\Phi=0$, where $\Phi$ represents
one of the auxiliary fields.
The stationary condition of $\Feff$ tells us that auxiliary fields
other than $\sigma$ and $\wt$
are expressed as functions of $(\sigma,\wt)$.
Thus the effective potential is a function of $T$, $\mu$,
and two order parameters; the chiral condensate $\sigma$ and $\wt$.
We can regard $\wt$ is a vector potential field for quarks;
the chemical potential shift is mainly determined by $\wt$,
and $\wt$ contributes repulsively to the effective potential in equilibrium.
This two order parameter feature may be a natural consequence
from the potential term from quarks, $\Vq(\mq;\tilmu,T)$. 
There are two independent derivatives,
$\partial\Vq/\partial\mq$ and $\partial\Vq/\partial\tilmu$,
which appear in the equilibrium condition,
then we have two degrees of freedom.

\section{Chiral Phase Transition in NNLO SC-LQCD}
\label{Sec:Results}

\begin{figure}[bt]
\begin{center}
\Psfig{8cm}{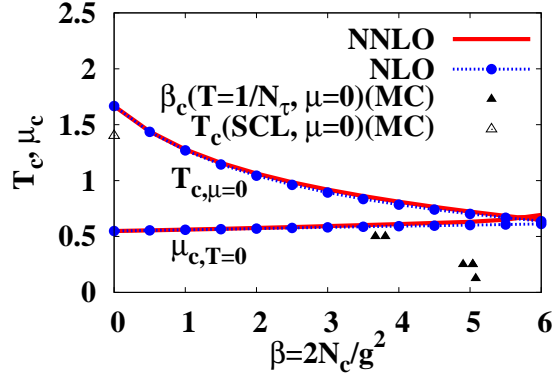}
\caption{The $\beta$ dependences
of a critical temperature at $\mu=0$ 
and a critical chemical potential at $T=0$.
We also show the MC results of $T_c$ in the strong coupling limit
and the critical coupling $\beta_c$ for a given $N_\tau$.
MC results are taken from~\protect{\cite{deForcrand:2009dh,MC}}.
}\label{Fig:Tc_muc}
\end{center}
\end{figure}

The effective potential determines the vacuum and the phase structure
of QCD matter.
The vacuum is determined by solving the stationary condition of $\Feff$,
which is equivalent to searching for the saddle point
of $\Feff$ in the $(\sigma,\wt)$ plane~\cite{PDevol}.
In Fig.~\ref{Fig:Tc_muc}, we show the 
the critical temperature at $\mu=0$ ($T_{c,\mu=0}$)
and the critical chemical potential at $T=0$ ($\mu_{c,T=0}$).
In both NLO and NNLO,
the phase transition on the $T$-axis is found to be the second order,
and $T_{c,\mu=0}$ decreases as $\beta$ increases.
NLO and NNLO give almost the same values of $T_{c,\mu=0}$.
We also show the MC results on $T_c$ in SCL
and the critical coupling ($\beta_c$)
for given values of $N_\tau=1/T$ at $\mu=0$
(filled triangles)~\cite{deForcrand:2009dh,MC}.
The suppression of $T_{c,\mu=0}$ in NLO and NNLO
is not enough to explain the MC results.
The phase transition on $\mu$-axis is numerically found to be the first-order
in NLO in the coupling range studied here ($\beta \leq 6$),
and for $\beta \lesssim 5.5$ in NNLO. 
The first order phase transition is determined by the magnitude relation
between $E_q$ and $\tilmu$.
As $\beta$ increases, both of $E_q$ and $\tilmu$ are suppressed,
and the suppression effects on $\tilmu$ is slightly larger than that in $E_q$.
As a result, $\mu_{c,T=0}^\mathrm{(1st)}$ is a slightly increasing function
of $\beta$.

\begin{figure}[bt]
\begin{center}
\Psfig{14cm}{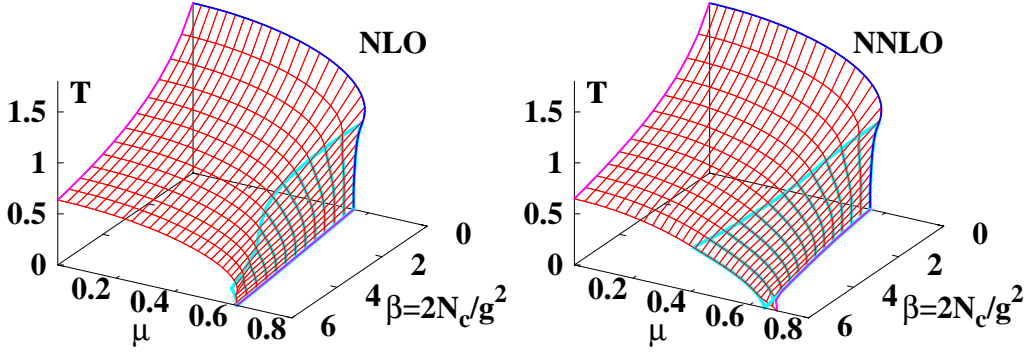}
\caption{
Phase diagram evolution in NLO and NNLO SC-LQCD.
Phase boundary are shown by thin line surface,
and thick surfaces show the first order phase transition boundary.
}\label{Fig:PDevolv}
\end{center}
\end{figure}

In Fig.~\ref{Fig:PDevolv}, we show the $\beta$ dependence 
of the phase diagram in NLO and NNLO.
While $T_{c,\mu=0}$ and $\mu_{c,T=0}$ are very similar in NLO and NNLO,
the shape of the phase boundary and the critical point 
are sensitive to the NNLO corrections.
In NLO,
the critical point (CP) temperature ($T_\mathrm{CP}$)
decreases and the CP chemical potential ($\mu_\mathrm{CP}$)
stays almost the same as $\beta$ increases.
The tri-critical point at strong coupling starts to deviate
from the second order phase boundary at $\beta \simeq 4.5$,
and it becomes the critical end point (CEP) even in the chiral limit.
In NNLO, both $T_\mathrm{CP}$ and $\mu_\mathrm{CP}$ decreases
as $\beta$ increases,
and the CP stays on the second order phase transition boundary.

The CP behavior in NNLO is considered to be an improvement over NLO.
Since one species of staggered fermion corresponds to $N_f=4$,
the phase transition at $\mu=0$ is expected to be the first order
in the continuum region due to anomaly contributions~\cite{Pisarski:1983ms}.
In the strong couping region, the chiral $\mathrm{SU}_f(4)$ symmetry 
is not manifest with staggered fermions.
The lattice QCD action with staggered fermion is invariant under
the chiral transformation, 
$\chi_x\to \exp(i\theta\epsilon_x)\chi_x$,
where $\epsilon_x=(-1)^{x_0+x_1+\cdots+x_d}$ is a $\gamma_5$
related phase~\cite{SCL}.
This transformation corresponds to the $\gamma_5\otimes\gamma_5$ rotation
in the spinor-flavor space in the continuum region,
which is only a small subgroup of $\mathrm{SU}_A(4)$.
As a result, effective number of flavors is small at strong coupling,
and the second order phase transition at $\mu=0$ emerges.
If we can take account of the finite coupling effects correctly,
the order of the phase transition at $\mu=0$ should change from the second 
to the first order at a certain coupling strength.
The behavior of CP in NNLO suggests that the first order transition boundary 
comes closer to the $T$-axis, and agrees with the above expectation
while the shift is not enough.

Another interesting point is the existence of the region
where $\mu_c^\mathrm{(1st)}(T)<\mu_c^\mathrm{(2nd)}(T)$ is satisfied.
Between these critical chemical potentials,
$\mu_c^\mathrm{(1st)}(T)<\mu<\mu_c^\mathrm{(2nd)}(T)$,
chiral condensate is suppressed compared to its vacuum value
but the chiral symmetry is not fully restored.
This partially chiral restored (PCR) matter suggested
in NLO SC-LQCD~\cite{Miura_QY} persists to exist also with NNLO corrections,
and would correspond to the quarkyonic matter suggested
at large $N_c$~\cite{QY}.

\section{Summary}
\label{Sec:Summary}
 
In this proceedings, we have derived 
an analytic expression of the effective potential
in the strong coupling lattice QCD (SC-LQCD)
with one species of unrooted staggered fermion for color $\mathrm{SU}(N_c)$
including the next-to-leading (NLO) and the next-to-next-to-leading (NNLO)
effects at finite temperature and density, 
and investigated finite coupling effects on the chiral phase transition
and the phase diagram.
Effective action terms have been systematically evaluated
based on the strong coupling expansion
in the leading order of the $1/d$ expansion.
NLO and NNLO effective action terms have been found to come
from one plaquette and two connected plaquette configurations.
We have applied the extended Hubbard-Stratonovich transformation 
to bosonize fermion interaction terms.
The next-to-nearest neighbor interaction appears in NNLO corrections,
and is evaluated by introducing a gluonic dressed fermion.
%
We have obtained the effective potential
as a function of temperature ($T$), chemical potential ($\mu$),
and the two order parameters:
the chiral condensate ($\sigma$) and the vector potential ($\wt$).
The vacuum is determined from the stationary condition
of the effective potential with respect to each auxiliary field.
NLO and NNLO effects result in 
modification of the wave function renormalization factor,
quark mass and chemical potential. 

The critical temperature at $\mu=0$ ($T_{c,\mu=0}$) 
and chemical potential at $T=0$ ($\mu_{c,T=0}$) 
are found to be similar in NLO and NNLO.
The critical temperature at $\mu=0$ ($T_{c,\mu=0}$)
is a decreasing function of $\beta$ in both NLO and NNLO,
while the suppression is not enough to explain the Monte-Carlo results
at $N_\tau = 1/T = 2, 4~\mathrm{and}~8$.
The critical chemical potential at $T=0$ ($\mu_{c,T=0}$)
is found to stay almost constant in both NLO and NNLO,
due to the cancellation of the effects from mass and effective chemical
potential reduction.
The shape of the phase boundary and the position of the critical point (CP)
are found to be sensitive to the NNLO effects.
As $\beta$ increases,
while CP moves towards the $\mu$-axis in NLO,
CP moves towards the $T$-axis in NNLO.

Inclusion of the Polyakov loop 
and higher order terms  in the $1/d$ expansion
would be interesting directions of study.

\section*{Acknowledgments}
We would like to thank
Philippe de Forcrand, Koichi Yazaki, Koji Hashimoto,
for useful discussions.
This work is supported in part by KAKENHI,
under the grant numbers,
17070002		
and 
19540252,		
the Global COE Program
"The Next Generation of Physics, Spun from Universality and Emergence",
and the Yukawa International Program for Quark-hadron Sciences (YIPQS).


\end{document}